%% file: sigconf.tex
\documentclass[sigconf]{acmart}
\AtBeginDocument{%
  }

\setcopyright{acmlicensed}
\copyrightyear{2018}
\acmYear{2018}
\acmDOI{XXXXXXX.XXXXXXX}
\acmConference[Conference acronym 'XX]{Make sure to enter the correct
  conference title from your rights confirmation email}{June 03--05,
  2018}{Woodstock, NY}
\acmISBN{978-1-4503-XXXX-X/2018/06}
\usepackage{graphicx}
\usepackage{eso-pic}

\begin{document}

\title{Sci-Surf: Navigating Scientific Literature Discovery through Human Feedback and Intelligent Summarization}

\author{Fang Guo}
\authornotemark[1] 
\affiliation{%
  \institution{Zhejiang University}
  \city{Hangzhou}
  \country{China}}
\email{guofang@westlake.edu.cn}

\author{Qi Zhu}
\authornote{Equal contribution.}
\affiliation{%
  \institution{Zhejiang University}
  \city{Hangzhou}
  \country{China}}
\email{qizhu.zju.research@gmail.com}

\author{Rongcan Pei}
\authornotemark[1] 
\affiliation{%
  \institution{Tongji University}
  \city{Shanghai}
  \country{China}}
\email{prc@tongji.edu.cn}

\author{Shuqi He}
\affiliation{%
  \institution{Google Cloud}
  \country{United States}}
\email{shuqihe95@gmail.com}

\author{Hui Chen}
\affiliation{%
  \institution{Zhejiang University}
  \city{Hangzhou}
  \country{China}}
\email{3220102841@zju.edu.cn}

\author{Yue Zhang}
\affiliation{%
  \institution{Westlake University}
  \city{Hangzhou}
  \country{China}}
\email{zhangyue@westlake.edu.cn}
\renewcommand{\shortauthors}{Guo et al.}


\begin{abstract}
The rapid growth of scientific publications makes it increasingly difficult for researchers to identify relevant new studies and effectively comprehend them. Existing academic discovery platforms typically rely on static topic subscriptions or embedding-based similarity and provide only abstracts or short summaries, offering limited support for nuanced intent modeling and in-depth paper summarization. We present \textbf{Sci-Surf}\footnote{
Project page: \href{http://www.sci-surf.com}{Sci-Surf Website}. 
Code and prompts: \href{https://github.com/Prongcan/sci-surf}{GitHub Repository}. 
Demo video: \href{https://drive.google.com/drive/folders/1fHltY5t2F11dycBbj3qXKL4wGrJdFZpP?usp=sharing}{Video Demo}.
}, an intent-centric knowledge discovery system that integrates feedback-driven personalized recommendation with multi-modal blog-style paper digestion.  Our approach refines user intent representations through LLM-based user profiling, while generating structured summaries that synthesize textual and visual information from full papers. The demo presents an end-to-end academic discovery pipeline and demonstrates measurable improvements in both recommendation quality and digestion quality through real-user evaluations. Specifically, the integration of verbalized profiles led to a \textbf{10.4\%} average improvement in predictive alignment with real-world user preferences throughout a month-long online evaluation.
\end{abstract}



\received{20 February 2007}
\received[revised]{12 March 2009}
\received[accepted]{5 June 2009}

\maketitle

\input{tex/1-intro}
\input{tex/2-problem}

\input{tex/3-exp}


\input{tex/6-concl}
\newpage

\bibliographystyle{ACM-Reference-Format}
\bibliography{sample-base}

\appendix

\end{document}

%% file: tex/1-intro.tex
\section{Introduction} 
\begin{figure*}[t!]
    \includegraphics[width=1\textwidth]{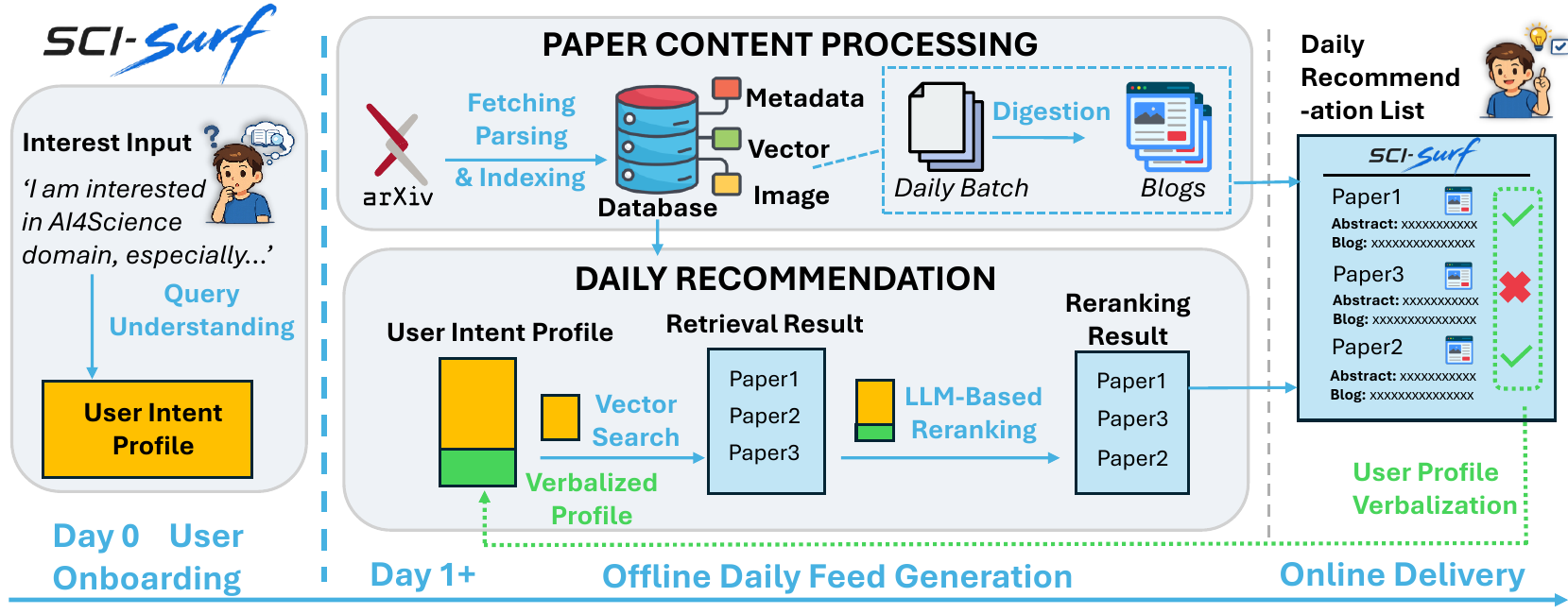}
    \caption{
    Overview of Sci-Surf. User intent is initialized during onboarding and continuously refined from interaction feedback, while newly published papers are processed into personalized recommendations and blog-style digests.
    }
    \label{pipeline}
\end{figure*}

The volume of publications in computer science is expanding at an unprecedented rate. For example, recent analyses~\cite{cunningham2025analysis} show that the number of annual CS journal publications has grown from roughly 20{,}000 in the year 2000 to over 120{,}000 in the year 2023, with the pace expected to accelerate further. This surge of scholarly output presents two fundamental challenges for modern academic discovery. First, researchers must identify publications that genuinely align with their research objectives amid an overwhelming stream of new work. Second, even when relevant papers are retrieved, effectively understanding their key contributions, methodologies, and implications requires substantial cognitive effort. Addressing these two challenges in a unified manner remains a central bottleneck in contemporary academic discovery systems.

Existing work on academic discovery primarily focuses on improving retrieval effectiveness. Embedding-based scientific retrieval models~\citep{Cohan2020SPECTERDR,cai2024mixgr,mandikal2024sparse,zhang2025scientific} and LLM-based techniques such as query rewriting, expansion, and agentic search~\cite{ma2023query,he2025pasa,shi2025spar,zhu2025large} improve query--document matching, while human-in-the-loop systems leverage active learning to reduce screening effort~\cite{van2021open,guo2023scimine}. However, most approaches operate in single-turn settings and do not maintain persistent representations of user intent. Similarly, deployed academic discovery platforms\footnote{\url{https://www.scholarinbox.com/}}\footnote{\url{https://www.semanticscholar.org/}}\footnote{\url{https://www.alphaxiv.org/}} provide interactive search experiences but largely optimize for immediate retrieval rather than continuously modeling and refining evolving user interests.

A second challenge lies in paper understanding. Existing platforms typically present abstracts or short TLDR-style summaries, which often omit methodological intuition, contextual background, and visual evidence essential for understanding modern scientific papers. As a result, even when relevant papers are successfully retrieved, researchers must still spend substantial effort digesting and interpreting them.

To address these challenges, we introduce \textbf{Sci-Surf}, an intent-centric academic discovery system that enables feedback-driven intent modeling with enhanced content presentation. It integrates two complementary components:  
(1) \textbf{feedback-aware reranking}, which iteratively verbalizes user intent profiles by leveraging large language models (LLMs) to reason over historical feedback; and  
(2) \textbf{multi-modal blog-style paper digestion}, which generates structured, extended summaries by synthesizing textual content and visual elements from the original paper into an accessible blog-style narrative.  

To support these capabilities in a real-world setting, Sci-Surf adopts a publication-synchronized, batch-based workflow (Figure~\ref{pipeline}). Newly released \textit{arXiv} papers are collected at fixed daily intervals and processed offline as a daily batch. LLM operations are executed during this pipeline to generate structured paper digests and daily personalized recommendations. This design shifts heavy computation away from user query time and decouples system processing from real-time user interaction.

Specifically, the system operates in two phases. 
On \textbf{Day 0}, users provide free-form descriptions of their research interests, which are normalized by an LLM into a structured \textbf{intent profile}. 
This intent profile serves as a long-term personalization anchor and is continuously refined based on user interactions. 
Beginning on \textbf{Day 1+}, Sci-Surf enters a daily feed generation pipeline consisting of three components: 
(1) \emph{Paper Content Processing}, which collects newly released papers into the daily batch and processes them for indexing and blog-style digestion; 
(2) \emph{Daily Recommendation}, which aligns the processed papers with the stored intent profile to generate daily recommendations; and 
(3) \emph{User Profile Verbalization}, which periodically analyzes accumulated user feedback (e.g., likes and dislikes) and updates the system’s internal representation of the intent profile, thereby influencing subsequent recommendation cycles.
In summary, Sci-Surf makes three contributions: (1) \textbf{Autonomous Intent Refinement}, which leverages LLM-based reasoning to continuously refine user intent profiles from interaction histories and bridge the gap between high-recall retrieval and long-term user interests; (2) \textbf{Structured Multimodal Digestion}, a blog-style paper understanding pipeline that synthesizes textual insights and visual evidence into accessible research digests; and (3) \textbf{Longitudinal Real-User Evaluation}, which validates both recommendation quality and digestion faithfulness through real-user studies, demonstrating the effectiveness of verbalized intent profiles in a dynamic academic discovery setting.

%% file: tex/2-problem.tex
\section{Sci-Surf Workflow}


Figure~\ref{pipeline} illustrates the Sci-Surf workflow.
The system consists of a user onboarding stage (Day 0) and a recurring daily feed generation stage (Day 1+).


\subsection{User Onboarding}

User onboarding occurs on Day 0 and initializes a persistent semantic intent profile. 
Users provide free-form descriptions of their research interests in natural language. 
The system normalizes this input into a structured intent profile through LLM-based semantic modeling.

The resulting intent profile serves as a long-term personalization anchor that guides the subsequent daily feed generation pipeline. 
In line with our batch-based design, recommendations are generated during the next-day scheduled cycle rather than at onboarding time, thereby avoiding real-time heavy computation and ensuring scalable, low-latency delivery.

\subsection{Daily Feed Generation}

Beginning on Day 1, Sci-Surf enters the publication-synchronized daily pipeline. Each day, newly published papers are aggregated into a daily batch, within which both content processing and intent matching are performed offline. As illustrated in Figure~\ref{pipeline}, the pipeline consists of three components: Paper Content Processing, Personalized Recommendation, and User Profile Verbalization.

\subsubsection{Paper Content Processing}

The paper content processing module collects newly released papers into the
daily batch and processes them for indexing and
blog-style digestion.

\textbf{Paper Fetching, Parsing, and Indexing.}
The workflow begins with the large-scale acquisition of newly released papers from the \textit{arXiv} Computer Science category, which receives 300--500 new submissions daily. 
To ensure robust ingestion, both HTML and PDF versions of each paper are fetched. 
The HTML version provides well-structured sections, captions, figures, and tables, while the PDF parser serves as a fallback to ensure high coverage when HTML content is incomplete or malformed. 
The PDF stream additionally enables vision-language model (VLM) based digestion, whose effectiveness is evaluated in the Experiments section.

Each paper is parsed into a structured representation containing textual chunks, visual elements (figures and tables), and mathematical expressions. 
These elements are stored across three coordinated components:
(1) a metadata database for bibliographic information (e.g., title, authors, publication date),
(2) a vector store (e.g., FAISS) populated with semantic embeddings derived from abstracts, and
(3) an object store for visual content that supports multi-modal digestion.



\textbf{Paper Digestion.}
As shown in Figure~\ref{pipeline}, once new papers are ingested, Sci-Surf enters the digestion stage, where each article is transformed into a structured, multi-modal blog-style digest.
Instead of generating summaries at recommendation time, the system precomputes digests for all newly batched papers, enabling low-latency access to high-quality overviews during ranking.

The summary prompt is carefully designed to ensure clarity, consistency, and interpretability. Each digest follows the sequence: \emph{TL;DR overview}, \emph{motivation}, \emph{contributions}, \emph{method summary}, \emph{analysis of figures/tables}, \emph{experimental results}, and \emph{future outlook}. This format gives readers both a high-level snapshot and a concise walkthrough of the paper's key elements.

Benefiting from rich elements generated during the previous content processing stage, we are able to inject all relevant elements, including image paths, tables, original text, and equations, into the prompt. We then instruct the LLM to insert image placeholders at appropriate positions. After the blog post is generated, we systematically replace all matched placeholders with the corresponding storage paths. By integrating these elements, the digest offers richer, more context-aware insight than conventional text-only summaries.

As a result, the digestion module converts raw full-text and multi-modal content into concise, high-quality summaries that support downstream personalized recommendations.

\subsubsection{Daily Recommendation}

This part of the pipeline delivers personalized and timely research updates based on each user’s stored intent profile.
As shown in Figure~\ref{pipeline}, the personalized recommendation module consists of two stages:
(1) a vector search stage, and
(2) an LLM-based reranking stage. 

\textbf{Vector Search.}
Using the structured user intent profile, the system performs embedding-based retrieval over the abstracts of newly ingested papers.
We adopt GritLM~\cite{Muennighoff2024GenerativeRI} as the base retriever due to its strong performance in our offline evaluation.
The intent profile is embedded and matched against paper embeddings stored in the vector database.

To ensure timeliness and reduce noise, retrieved candidates are further refined through:
(i) a freshness constraint prioritizing submissions from the past 3--5 days,
(ii) de-duplication to prevent resurfacing previously recommended items, and
(iii) similarity thresholding to remove weak semantic matches.
This stage produces a relevance-ranked candidate set for subsequent reranking.

\textbf{LLM-based Reranking.}
Although vector retrieval provides efficient semantic matching, it may not fully capture nuanced contextual relevance.
Therefore, the top candidates are passed to an LLM-based reranking stage.
In this stage, the LLM evaluates the alignment between each paper’s pre-generated digest and the user’s structured intent profile, enabling deeper reasoning beyond embedding similarity.

The reranker refines the ordering of candidates by considering finer-grained factors such as methodological focus, problem setting, and application domain.
This second-stage reasoning improves precision and ensures that recommended papers better reflect the user’s research trajectory.


The reranked top-$k$ papers are finalized as the daily recommendation list and delivered to the user. Subsequent user interactions with these recommendations are logged and incorporated into the profile verbalization process in future recommendation cycles. In our demo, we set $k=5$.

\subsubsection{Personalization via User Profile Verbalization}

To support longitudinal personalization, Sci-Surf features a feedback-driven user intent profile verbalization component that self-improves recommendation quality along the user journey. 

Given a user $u$ with historical interactions over $N$ days, we define:

\begin{itemize}
    \item $\mathcal{D} = \{d_1, d_2, \ldots, d_N\}$: Set of days with positive feedback
    \item $\mathcal{P}_i = \{p_{i,1}, p_{i,2}, \ldots, p_{i,k_i}\}$: Set of recommended papers for day $d_i$
    \item $\mathcal{Y}_i \subseteq \mathcal{P}_i$: Set of positively labeled papers (liked) for day $d_i$
    \item $q_i$: Query/research interest for day $d_i$
\end{itemize}

We split days chronologically into 70\% training $\mathcal{D}_{train}$ and 30\% validation days $\mathcal{D}_{val}$.

For each training day $d_i \in \mathcal{D}_{train}$, we construct a list of labeled examples $\{p,y\}$, where $y \in \{0,1\}$ is the positive or null feedback received from the user. 

\subsection*{Profile Extraction}

We use an LLM-based Verbalizer $\mathcal{V}$ to extract user preferences from PDF content:
\begin{equation}
    \pi_u = \mathcal{V}\left(\mathcal{T}, \{\text{PDF}(p) : (p, \cdot) \in \mathcal{T}\}\right)
\end{equation}
where $\pi_u$ denotes the user profile, consisting of a persona description $\pi_u^{persona}$, a set of negative constraints $\pi_u^{neg}=\{c_1,\ldots,c_m\}$, and a set of ranking heuristics $\pi_u^{heur}=\{h_1,\ldots,h_l\}$.


We utilize the validation set $\mathcal{D}_{val}$ to improve the verbalization prompt. Once the user profile has been extracted, it will be instantiated to the LLM reranker prompt in addition to the interest description the user provided. 

The complete user profile verbalization prompt and personalized ranking prompt are provided in our github repository for reference. In practice, we find that the iterative updating profile as part of the ranking prompt, instead of fully rewriting, has the most robust behavior across different users.





%% file: tex/3-exp.tex
\section{Experiments}




We evaluate Sci-Surf from two complementary perspectives: (1) \textbf{Personalized Recommendation Effectiveness}, covering both retrieval and reranking performance, and (2) \textbf{Digestion Faithfulness}, assessing the factual consistency of generated blog-style digests through automatic and human evaluation.

\subsection{Personalized Recommendation Effectiveness}
\subsubsection{Offline Retrieval Benchmark}


\begin{table}[h]
\centering
\small
\caption{Retrieval performance on the LitSearch benchmark.}
\begin{tabular}{lcc}
\toprule
\textbf{Method} & \textbf{Recall@5} & \textbf{Recall@20} \\
\midrule
BM25      & 0.522 & 0.673 \\
SPECTER   & 0.515 & 0.668 \\
GritLM    & 0.705 & 0.823 \\
BGE-M3    & 0.673 & 0.802 \\
Hybrid    & 0.682 & 0.811 \\
\bottomrule
\end{tabular}
\label{tab:litsearch}
\end{table}


We evaluate retrieval quality on \textit{LitSearch}, a benchmark containing 597 scientific queries and expert relevance judgments. We compare BM25, three embedding-based retrievers: SPECTER, GritLM, and BGE-M3~\cite{Chen2024BGEMM}, and a hybrid BM25+embedding baseline using Reciprocal Rank Fusion (RRF).

Table~\ref{tab:litsearch} reports Recall@5 and Recall@20. Embedding-based retrieval consistently outperforms BM25, with GritLM achieving the best performance: 0.705 for Recall@5 and 0.823 for Recall@20. Hybrid fusion provides limited additional benefit, suggesting that lexical signals contribute little beyond strong semantic retrieval. We therefore adopt GritLM as the default retriever in Sci-Surf.

\subsubsection{Reranking Evaluation with Real-User Interaction Logs}

\begin{table}[h]
\centering
\small
\caption{Evaluation of personalized recommendations on 15 real users. ``+ Profile'' denotes the inclusion of verbalized user profiles in the reranking prompt.}
\label{tab:personalization}

\begin{tabular}{lccc}
\toprule
\textbf{Setting} & \textbf{Rel.\%} & \textbf{Very Rel.} & \textbf{Rel.} \\
\midrule
Baseline   & 18.9\% & 0.5\% & 18.4\% \\
+ Profile  & \textbf{29.3\%} & \textbf{3.6\%} & \textbf{25.7\%} \\
\bottomrule
\end{tabular}

\end{table}

To evaluate the effectiveness of the LLM-based reranking module in a realistic deployment setting, we use interaction logs from 15 active Sci-Surf users collected between November 2025 and May 2026 to construct verbalized user profiles. We perform an ablation study comparing ranking outputs generated with and without user profiles. For each user, papers appearing exclusively in either the personalized or baseline top-5 recommendation list are manually annotated in a blind evaluation.

Annotators label each paper as \textit{highly relevant}, \textit{domain-relevant}, or \textit{irrelevant}. We report \textit{Very Rel.}, \textit{Rel.}, and \textit{Rel.\%}, corresponding to the proportions of highly relevant papers, domain-relevant papers, and their combined proportion, respectively.

Table~\ref{tab:personalization} summarizes the aggregated results. Incorporating verbalized user profiles increases overall relevance (\textit{Rel.\%}) from 18.9\% to 29.3\% and improves the proportion of highly relevant papers from 0.5\% to 3.6\%. These results indicate that profile verbalization effectively captures fine-grained research preferences beyond generic retrieval signals, leading to recommendations that are more closely aligned with users' current interests.

\subsection{Digestion Faithfulness}

\subsubsection{Automated Hallucination Detection}

To evaluate the factual faithfulness of blog-style digests, we conduct a hallucination detection study on 1,000 paper--blog pairs using DeepSeek-Reasoner. Given the full paper text and the corresponding digest, the model identifies unsupported or inconsistent statements and categorizes them as either \textit{critical errors} (substantial misrepresentations) or \textit{minor issues} (e.g., imprecise phrasing or incomplete grounding). To balance performance and cost, we compare Gemini-2.5-Flash with Qwen-32B and Qwen-235B-A22B using HTML-parsed, JSON-formatted inputs.

Overall, Gemini substantially outperforms both Qwen models in factual reliability. The JSON-based Gemini pipeline achieves the lowest hallucination rate, averaging 0.22 critical and 0.76 minor errors per digest, compared with over 2 critical and 1.8 minor errors for both Qwen variants. PDF-based Gemini remains competitive (0.43/0.86 critical/minor errors) but is consistently outperformed by structured JSON inputs, suggesting that cleaner representations improve factual grounding and reduce error propagation during digest generation.

\subsubsection{Human Evaluation Study}

To verify the reliability of the LLM-based detector, we conduct a human evaluation study measuring alignment between model and human judgments. 
We recruit three users and sample 30 recommended papers per user (90 digests in total). 
Annotators evaluate each digest following the same rubric as the automated detector.

Agreement is measured with Pearson correlation and Cohen’s $\kappa$. For overall faithfulness, the Pearson correlation reaches 0.61, and Cohen’s $\kappa$ is 0.45, indicating moderate-to-strong consistency.
\looseness=-1
These findings suggest that the proposed LLM-based detector serves as a reasonably reliable proxy for human evaluation, providing additional evidence that the generated blog-style digests maintain a satisfactory level of factual faithfulness to the original papers.




%% file: tex/6-concl.tex
\section{Conclusion}
Sci-Surf is an intent-centric academic discovery system that integrates feedback-driven recommendation with multi-modal blog-style paper digestion. 
By leveraging LLM-based user profiling to maintain evolving intent representations and generating structured, faithful digests from full papers, the system supports both nuanced preference modeling and in-depth content comprehension. 
Moving forward, we plan to incorporate richer feedback modeling and conduct larger-scale user studies to further improve intent understanding.

%% file: sample-base.bib
@article{cunningham2025analysis,
  title={An analysis of the impact of gold open access publications in computer science},
  author={Cunningham, P{\'a}draig and Smyth, Barry},
  journal={Communications of the ACM},
  volume={68},
  number={8},
  pages={62--69},
  year={2025},
  publisher={ACM New York, NY, USA}
}

@article{Cohan2020SPECTERDR,
  title={SPECTER: Document-level Representation Learning using Citation-informed Transformers},
  author={Arman Cohan and Sergey Feldman and Iz Beltagy and Doug Downey and Daniel S. Weld},
  journal={ArXiv},
  year={2020},
  volume={abs/2004.07180},
  url={https://api.semanticscholar.org/CorpusID:215768677}
}

@inproceedings{Chen2024BGEMM,
  title={BGE M3-Embedding: Multi-Lingual, Multi-Functionality, Multi-Granularity Text Embeddings Through Self-Knowledge Distillation},
  author={Jianlv Chen and Shitao Xiao and Peitian Zhang and Kun Luo and Defu Lian and Zheng Liu},
  booktitle={Annual Meeting of the Association for Computational Linguistics},
  year={2024},
  url={https://api.semanticscholar.org/CorpusID:267413218}
}

@article{Muennighoff2024GenerativeRI,
  title={Generative Representational Instruction Tuning},
  author={Niklas Muennighoff and Hongjin Su and Liang Wang and Nan Yang and Furu Wei and Tao Yu and Amanpreet Singh and Douwe Kiela},
  journal={ArXiv},
  year={2024},
  volume={abs/2402.09906},
  url={https://api.semanticscholar.org/CorpusID:267681873}
}

@article{zhang2025scientific,
  title={Scientific paper retrieval with llm-guided semantic-based ranking},
  author={Zhang, Yunyi and Yang, Ruozhen and Jiao, Siqi and Kang, SeongKu and Han, Jiawei},
  journal={arXiv preprint arXiv:2505.21815},
  year={2025}
}

@article{mandikal2024sparse,
  title={Sparse meets dense: A hybrid approach to enhance scientific document retrieval},
  author={Mandikal, Priyanka and Mooney, Raymond},
  journal={arXiv preprint arXiv:2401.04055},
  year={2024}
}

@inproceedings{cai2024mixgr,
  title={MixGR: Enhancing retriever generalization for scientific domain through complementary granularity},
  author={Cai, Fengyu and Zhao, Xinran and Chen, Tong and Chen, Sihao and Zhang, Hongming and Gurevych, Iryna and Koeppl, Heinz},
  booktitle={Proceedings of the 2024 Conference on Empirical Methods in Natural Language Processing},
  pages={10369--10391},
  year={2024}
}

@article{zhu2025large,
  title={Large language models for information retrieval: A survey},
  author={Zhu, Yutao and Yuan, Huaying and Wang, Shuting and Liu, Jiongnan and Liu, Wenhan and Deng, Chenlong and Chen, Haonan and Liu, Zheng and Dou, Zhicheng and Wen, Ji-Rong},
  journal={ACM Transactions on Information Systems},
  volume={44},
  number={1},
  pages={1--54},
  year={2025},
  publisher={ACM New York, NY}
}

@inproceedings{ma2023query,
  title={Query rewriting in retrieval-augmented large language models},
  author={Ma, Xinbei and Gong, Yeyun and He, Pengcheng and Zhao, Hai and Duan, Nan},
  booktitle={Proceedings of the 2023 Conference on Empirical Methods in Natural Language Processing},
  pages={5303--5315},
  year={2023}
}

@inproceedings{he2025pasa,
  title={Pasa: An llm agent for comprehensive academic paper search},
  author={He, Yichen and Huang, Guanhua and Feng, Peiyuan and Lin, Yuan and Zhang, Yuchen and Li, Hang and others},
  booktitle={Proceedings of the 63rd Annual Meeting of the Association for Computational Linguistics (Volume 1: Long Papers)},
  pages={11663--11679},
  year={2025}
}

@article{shi2025spar,
  title={Spar: Scholar paper retrieval with llm-based agents for enhanced academic search},
  author={Shi, Xiaofeng and Li, Yuduo and Kou, Qian and Yu, Longbin and Xie, Jinxin and Zhou, Hua},
  journal={arXiv preprint arXiv:2507.15245},
  year={2025}
}

@inproceedings{guo2023scimine,
  title={Scimine: An efficient systematic prioritization model based on richer semantic information},
  author={Guo, Fang and Luo, Yun and Yang, Linyi and Zhang, Yue},
  booktitle={Proceedings of the 46th International ACM SIGIR Conference on Research and Development in Information Retrieval},
  pages={205--215},
  year={2023}
}

@article{van2021open,
  title={An open source machine learning framework for efficient and transparent systematic reviews},
  author={Van De Schoot, Rens and De Bruin, Jonathan and Schram, Raoul and Zahedi, Parisa and De Boer, Jan and Weijdema, Felix and Kramer, Bianca and Huijts, Martijn and Hoogerwerf, Maarten and Ferdinands, Gerbrich and others},
  journal={Nature machine intelligence},
  volume={3},
  number={2},
  pages={125--133},
  year={2021},
  publisher={Nature Publishing Group UK London}
}
